\documentclass[aps,prb,twocolumn,superscriptaddress,floatfix,bibliography]{revtex4-2}
\usepackage{amsfonts,amsmath,amsthm,amssymb,amsopn,graphicx}
\usepackage{dsfont,bbm,xcolor}
\usepackage{bm}
\usepackage{enumitem}
\usepackage{hyperref}
\usepackage{breqn}
\usepackage{siunitx}
\usepackage{color}
\usepackage[]{breqn}
\usepackage{mathtools}
\usepackage{bigints}
\usepackage[caption=false]{subfig}

\makeatletter
\let\cat@comma@active\@empty
\makeatother

\def\bra#1{\mathinner{\langle{#1}|}}
\def\ket#1{\mathinner{|{#1}\rangle}}
\def\abs#1{\left | #1 \right |}
\newcommand{\braket}[2]{\langle #1 \vert #2 \rangle}
\newcommand{\ketbra}[2]{\vert  #1\rangle   \langle #2 \vert}

\def\bra#1{\mathinner{\langle{#1}|}}
\def\ket#1{\mathinner{|{#1}\rangle}}

\newcommand{\intbz}{\int_{\rm{BZ}}d\textbf{k}}

\newcommand{\mode}{\ \ (\textrm{mod}\ e)}

\usepackage{amsthm}
\theoremstyle{definition}

\begin{document}
\title{Microscopic theory of Chern polarization via crystalline defect charge}
\author{Thivan M. Gunawardana}
\affiliation{Department of Mathematics, Imperial College London, London SW7 2AZ, United Kingdom}
\author{Frank Schindler}
\affiliation{Blackett Laboratory, Imperial College London, London SW7 2AZ, United Kingdom}
\author{Ari M. Turner}
\affiliation{Department of Physics, Technion, Haifa 320000, Israel}
\author{Ryan Barnett}
\affiliation{Department of Mathematics, Imperial College London, London SW7 2AZ, United Kingdom}
\begin{abstract}
    The modern theory of polarization does not apply in its original form to systems with non-trivial band topology. Chern insulators are one such example. Defining polarization for them is complicated because they are insulating in the bulk but exhibit metallic edge states. Wannier functions formed a key ingredient of the original modern theory of polarization, but it has been considered that these cannot be applied to Chern insulators since they are no longer exponentially localized and the Wannier center, obtained from the Zak phase, is no longer gauge invariant. In this article, we provide an unambiguous definition of absolute polarization for a Chern insulator in terms of the Zak phase. We obtain our expression by studying the non-quantized fractional charge bound to lattice dislocations. Our expression can be computed directly from bulk quantities and makes no assumption on the edge state filling. It is fully consistent with previous results on the quantized charge bound to dislocations in the presence of crystalline symmetry. At the same time, our result is more general since it also applies to Chern insulators which do not have crystalline symmetries other than translations.
\end{abstract}
\maketitle

\section{Introduction} The modern theory of polarization \cite{moderntheory, surface93} marked a major step in the understanding of the microscopic behavior of electrons in crystals. By treating changes in polarization as the fundamental physical quantity, this theory allows for the extension of a notion of dipole moment per unit volume to periodic crystals via a Berry phase formalism. This is particularly non-trivial because the electrons do not behave as point charges but are rather governed by wavefunctions. A major consequence of this result is that changes in polarization can be cast in terms of Wannier functions \cite{wannier37, wannier12}, with the electronic charge centers being simply given by the Wannier centers. Moreover, the surface charge theorem \cite{surface93} allows for the sensible definition of an {\it absolute} polarization also in terms of the Wannier centers. This theory assumes that the system is insulating throughout and that globally smooth  Bloch states can be constructed. However, there are systems with non-trivial band topology for which these assumptions do not hold. Chern insulators, characterized by an integer topological invariant known as the Chern number, are the canonical example of such a topological band system. For such systems, there is a difficulty with defining the polarization via the surface charge theorem because the charge density on the surface can be varied arbitrarily by changing the occupation of the edge states. So the question we will ask is: ``Is there a meaning to polarization in a Chern insulator?"

The extension of the modern theory of polarization to Chern insulators has  received a considerable amount of attention in recent years~\cite{coh09,barkeshli23,Mahon23,Vaidya24,zhang2024}. In Ref.~\cite{coh09}, it was shown that the change in polarization during an adiabatic deformation could be computed for a Chern insulator without the use of Wannier functions. Moreover, the surface charge theorem \cite{surface93} was also extended but was shown to depend on the filling of metallic edge states, where these edge states were treated on a separate footing. More recently, Ref. \cite{Vaidya24} showed how bound charge at an interface between two Chern insulators with the same Chern number could be used to identify a polarization difference between them, focusing mainly on systems with inversion symmetry. Since Ref. \cite{coh09}, much progress has also been made on defining an absolute polarization for topological crystalline insulators (TCIs) \cite{tci11}. These systems possess additional crystalline symmetries beyond translations, such as inversion symmetry, which means that a sensible definition of polarization must be quantized to high symmetry points in the lattice. A straightforward way to probe absolute polarization in these systems is via fractional charges bound to lattice defects \cite{teo10, juricic12}. For TCIs without an exponentially localized and symmetry-preserving Wannier representation, charges at lattice dislocations can still be written in terms of weak topological invariants based on eigenvalues of crystalline symmetry operators \cite{vanmiert18, schindler22}. Charges bound to lattice disclinations can be found in a similar fashion \cite{hughes20} and this has been crucial to the study of the recently introduced higher order topological insulators~\cite{schindler18, benalcazar19}, which are insulators that do not possess gapless edge states but can instead have topologically protected corner states.
In the recent Refs.~\cite{barkeshli23} and \cite{barkeshli24}, a topological field theoretic approach was used to define absolute polarization for TCIs with non-trivial Chern number via the aforementioned bound charges at lattice dislocations and disclinations.

In this article, we present a more direct and microscopic approach to defining and evaluating absolute polarization for a Chern insulator via 
the Zak phase \cite{zak}, so that it can be expressed explicitly in terms of the bulk wavefunctions of the electrons. In previous work~\cite{ourpaper}, we studied changes in polarization under adiabatic deformations within a Wannier state picture and obtained results consistent with Ref.~\cite{coh09}.
In this article, we determine charge bound to crystalline lattice defects in general Chern insulators, requiring only bulk translational invariance. Using this, we are able to determine a simple gauge-invariant expression for the absolute polarization that is a bulk quantity. The expressions for relative polarization \cite{ourpaper} and absolute polarization are connected in a very simple way which will be explained. While our result is consistent with those found in TCIs, it has the advantage of not relying on crystalline symmetries other than bulk translational invariance and is in this sense more general. For such systems, the dislocation charge can take on a continuous range of values as Hamiltonian parameters are varied, in contrast to TCIs involving extra crystalline symmetries.

The paper is organized as follows. In Section II, we set the notation and, to be self-contained, provide known results on surface charge and polarization for 1D and 2D systems. Key formulas that will be used later are derived. Section III contains the main results of the work. We present a derivation of the bound charge found on a lattice dislocation for a general Chern insulator. In Section IV, we further demonstrate the validity of our result by verifying it for a paradigmatic tight-binding model of a Chern insulator. In Section V, we discuss and conclude. 

\section{Background}

The purpose of this section is to set notation and to motivate known results in the context of the present problem that will be needed later.

\subsection{Setup and notation}
In this article, we shall work exclusively within the tight-binding framework. Extending our results to continuum models should not pose any difficulties. With this in mind and assuming that we always have translational invariance in the bulk, we use a basis composed of states of the form $\ket{\textbf{R}} \otimes \ket{\alpha}$, where $\textbf{R}$ is the lattice vector of the unit cell, and $\alpha$ describes the degrees of freedom within a unit cell. For simplicity, we will use a 2D square lattice geometry with lattice constant $a$ throughout this work, but the extension to other 2D lattice geometries should be straightforward.

Within the tight-binding framework, there are two conventions for the choice of position operator. We shall use the position operator which treats all intra-cell degrees of freedom as being placed at the origin of each unit cell. In other words, we use the position operator given by
\begin{equation}
    \hat{\textbf{r}} = \sum_{\textbf{R}} \textbf{R}\ket{\textbf{R}}\bra{\textbf{R}} \otimes \mathds{1}.
    \label{position}
\end{equation}
We make this choice for simplicity; a position operator with knowledge of any crystal basis could be used instead, but this requires information not explicitly contained within the tight binding model. In the context of changes of polarization, the two different choices correspond to different ways of measuring the cell-averaged adiabatic current. A good discussion of this is given in Ref.~\cite{watanabe18}.

The aim of this work is to present a microscopic definition of polarization for a Chern insulator which is consistent with the trivial (Chern number zero) case. The polarization for a trivial 2D insulator can be written in terms of separate ionic and electronic contributions as
${\bf P}= {\bf P}_{\textrm{e}} + {\bf P}_{\rm ion}$.
For a single band, the electric polarization as given by the modern theory of polarization \cite{moderntheory} is
\begin{equation}
\label{Pe}
    \textbf{P}_{\textrm{e}} = -\frac{e\overline{\textbf{r}}}{V_{c}}\quad  \left(\text{mod}\ \frac{e\textbf{R}}{V_c}\right),
\end{equation}
where $e>0$ is the charge quantum, $\overline{\textbf{r}} = (\overline{x}, \overline{y})$ is the bulk Wannier center within the unit cell, $\textbf{R}$ is a real-space lattice vector and $V_c$ is the area of the real-space unit cell. Note that $\textbf{P}_{\textrm{e}}$ is gauge invariant up to a polarization quantum for a trivial insulator, for which exponentially localized Wannier functions can be constructed. The bulk Wannier center can be expressed in terms of the Bloch states as
\begin{equation}
\label{wc}
    \overline{\textbf{r}} = \frac{V_c}{(2\pi)^2}\intbz\ \textbf{A}(\textbf{k}),
\end{equation}
where $\textbf{A}(\textbf{k}) = i\bra{u_{\textbf{k}}}\nabla_{\textbf{k}}\ket{u_{\textbf{k}}}$ is the Berry connection of the band in terms of the cell-periodic parts of the Bloch states. In the rest of this work we will drop the ionic contribution to the polarization, taking the ionic charges to sit on sites of the Bravais lattice (i.e the ions are placed at points corresponding to real space lattice vectors $\textbf{R}$). Reincorporating it, if needed, is a straightforward exercise.


\subsection{Surface charge}
\label{surfcharge}
One of the major places where a notion of absolute polarization plays a role in trivial insulators is the surface charge theorem \cite{surface93}. In order to define absolute polarization for Chern insulators, it is natural to investigate whether this theorem can be extended to a Chern insulator. The surface charge theorem for a trivial 2D insulator relates the polarization defined by (\ref{Pe}) to the  bound charge per unit length on an edge of the system as
\begin{equation}
\label{scthm}
    \sigma = \textbf{P}\cdot\textbf{n}\ \left(\textrm{mod}\ \frac{e}{a_{||}}\right),
\end{equation}
where $\textbf{n}$ is the unit normal perpendicular to the edge and $a_{||}$ is the lattice constant in the direction parallel to the edge. This theorem assumes that the edge has the same translation symmetry as the bulk, and that it is insulating. In fact, this relation can be viewed as a way of defining the absolute polarization as it is uniquely fixed (modulo polarization quanta) by the bound charges on the edges of a conventional insulator. We can also use this to define polarization when there are non-topological metallic edge states, in which case the edge can theoretically be doped so that it becomes insulating. It is remarkable that these edge charges are determined by bulk properties \cite{surface93} for the conventional case.

On the other hand, matters become less straightforward if the edges are metallic in a topological sense (i.e they cannot be doped to make them insulating), like for the case of Chern insulators. (From this point onwards, we shall mean such edge states whenever we refer to metallic edge states.) How these metallic edges are populated will certainly influence the surface charge.  The method of edge termination or weak disorder will influence the edge filling, and therefore the polarization if defined using this logic. In the remainder of this background section we will motivate the surface charge theorem in 1D and 2D, highlighting difficulties encountered for a Chern insulator. Some of the results derived in this section will also be crucial to the derivation of the main result of this article.

\subsubsection{1D surface charge theorem}
We will now motivate the classic 1D surface charge theorem within the context of a tight-binding model. The main result of this subsection will be used extensively throughout the manuscript.
Consider a finite 1D system, with lattice constant $a$, which is periodic in the bulk with the ions placed at the origin of each unit cell. Each unit cell in the bulk is charge neutral in the ground state. Thus, given our convention for the tight-binding position operator (\ref{position}), it follows that the total charge density in the ground state (including the ionic contribution) vanishes in the bulk. If instead the position operator were to include intracell positions of the lattice basis, then a cell-averaged charge density would have to be used instead. We now place a partition in our system at a unit cell boundary somewhere deep in the bulk, dividing the system into a left part and a right part. We can then write the charge at the right edge of our 1D system as
\begin{equation}
    Q_R^{\textrm{1D}} = -e\textrm{Tr}(PS_R) + e\mathcal{N}_R^{\textrm{ion}},
    \label{1dcharge}
\end{equation}
where $P$ projects onto the single-particle states that are occupied in the many-body ground state of the insulating system, $S_R$ is a spatial projector which projects onto the right side of our partition and $\mathcal{N}_R^{\textrm{ion}}$ is the number of ions on the right side of the partition. We use a set of coordinates such that the unit cell immediately to the right of the partition is the ``home" unit cell (that is, the origin of that unit cell is taken to be the origin of the coordinate system).

\begin{figure}
    \centering
    \includegraphics[width = \columnwidth]{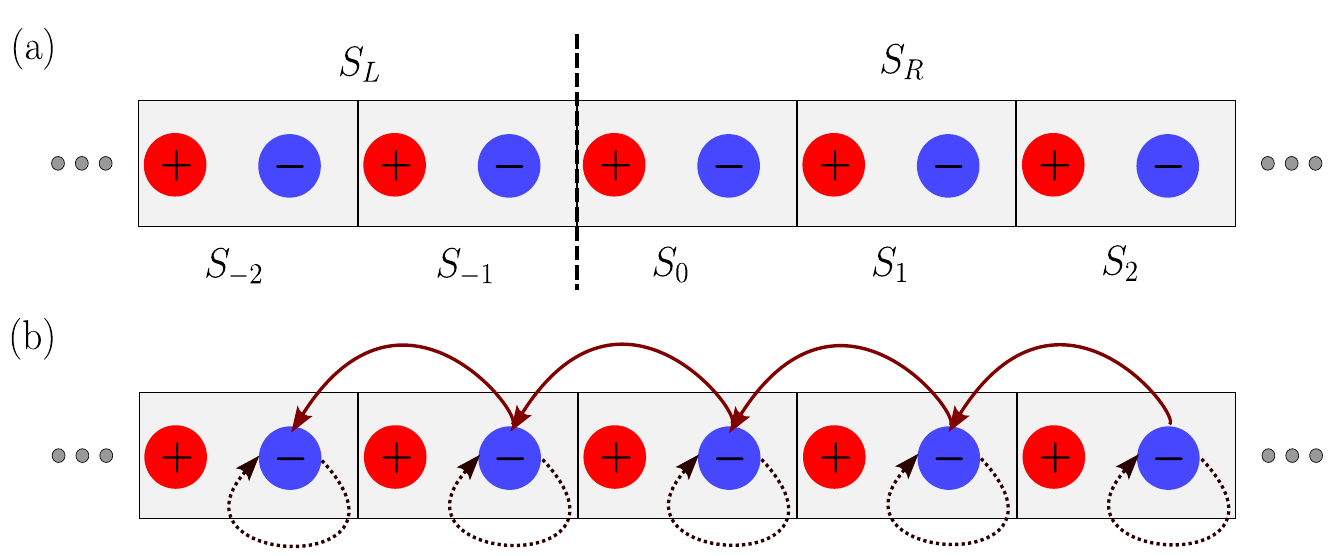}
    \caption{(a) Partitioning a 1D system deep in the bulk for the edge charge calculation. The ions are shown in red while the electrons (in the form of their Wannier centers) are shown in blue. The vertical dashed line denotes the partition. The regions into which the left and right spatial projectors, as well as individual cell projectors, project are shown. (b) Schematic of the flow of the HWF center of the effective 1D system as $k_y$ is taken from $0$ to $2\pi/a$. The solid (brown) arrows depict the non-trivial winding of the Wannier center for a Chern insulator, while the dotted (black) arrows depict the flow in the trivial case.}
    \label{wannier_winding}
\end{figure}

Next, we can construct an \textit{exponentially} localized (Wannier-like) basis for the occupied subspace of our finite system using the eigenstates of the projected position operator $PxP$ \cite{kohn73,marzari97}, where the position operator $x$ is given by the $x$-component of (\ref{position}). Deep in the bulk, these correspond to the maximally localized Wannier functions constructed using Bloch states for the periodic case. Denoting the basis state localized to the $j$th unit cell by $\ket{w_j}$, and its corresponding projector by $W_j = \ketbra{w_j}{w_j}$, the projector $P$ can now be written in this localized basis as $P = W_L + W_R$, where we have separated the left and right contributions $W_L = \sum_{j \in L} W_j$ and $W_R = \sum_{j \in R} W_j$. Similarly, the spatial projectors onto the left and right of the partition can be written as $S_L = \sum_{j \in L} S_j$ and $S_R = \sum_{j \in R} S_j$, where $S_j$ projects onto the $j$th unit cell. Fig.~\ref{wannier_winding}(a) shows a schematic of a 1D system partitioned into two pieces, with the various spatial projections around the partition labeled.

Returning to (\ref{1dcharge}), we can then expand the trace as follows:
\begin{equation}
    \textrm{Tr}(PS_R) = \textrm{Tr}(W_LS_R) - \textrm{Tr}(W_RS_L) + \textrm{Tr}(W_R),
\end{equation}
where we have used the fact that $S_R + S_L = \mathds{1}$. Note that
\begin{align}
    &\textrm{Tr}(W_LS_R) = \textrm{Tr}(W_{-1}S_0) + \left(\textrm{Tr}(W_{-2}S_0) + \textrm{Tr}(W_{-1}S_1)\right) \nonumber \\
    &\ + \left(\textrm{Tr}(W_{-3}S_0) + \textrm{Tr}(W_{-2}S_1) + \textrm{Tr}(W_{-1}S_2)\right) + ...
    \label{real_space_expansion}
\end{align}
Given the exponential localization of the $\ket{w_j}$'s, the sets of terms in the expansion (\ref{real_space_expansion}) decay exponentially. This is because we are computing the densities of our localized states to the left of the partition at sites on the right which are further and further away. Thus, only those contributions local to the partition will be non-trivial. Such contributions are purely from the bulk. A similar argument applies to $\textrm{Tr}(W_RS_L)$. Using the translational invariance of the bulk, we can therefore write
\begin{align}
    \textrm{Tr}(PS_R) &= \sum_n n\textrm{Tr}(W_0S_n) + \textrm{Tr}(W_R) \nonumber \\
    &= \frac{1}{a}\bra{w_0}x\ket{w_0} + \mathcal{N}^{\textrm{e}}_R,
\end{align}
where $\mathcal{N}^{\textrm{e}}_R$ is the (integer) number of localized electronic charge centers on the right of the partition.

Finally, noting that we can take $\bra{w_0}x\ket{w_0}$ to be the bulk Wannier centre (since the home unit cell is deep in the bulk), we arrive at the celebrated 1D surface charge theorem
\begin{equation}
    Q_R^{\textrm{1D}} = -\frac{e}{a}\bra{w_0}x\ket{w_0} + e(\mathcal{N}^{\textrm{ion}}_R - \mathcal{N}^{\textrm{e}}_R).
    \label{1dsurfcharge}
\end{equation}
The second term in (\ref{1dsurfcharge}) is an integer multiple of $e$, which counts the total number of ions and Wannier centers to the right of the partition. The first term, which is the bulk dipole moment per unit length, gives a fractional charge at the edge of the system when the bulk polarization is non-zero. It should be emphasized that this dipole moment is a bulk quantity whilst the second term in (\ref{1dsurfcharge}), though sensitive to edge details, is restricted to be an integer multiple of $e$. Therefore the 1D edge charge (up to a charge quantum) is determined entirely by bulk properties, even if there are surface irregularities. The above can be directly compared with ($\ref{Pe}$) and ($\ref{scthm}$).

\subsubsection{2D surface charge }

We now consider extending
the 1D surface charge theorem to the 2D case through the process of dimensional reduction \cite{Qi08}. Recalling that we are considering a square lattice with lattice constant $a$, we take the system to be periodic along the $y$-direction and open along the $x$-direction, corresponding to a cylindrical geometry. Then, $k_y$ remains a good quantum number and we have an effective 1D system for each value of $k_y$. That is, we can take an inverse Fourier transform of the bulk momentum space Hamiltonian in just the $x$-direction in order to obtain an effective real-space 1D Hamiltonian $H(k_y)$ for each $k_y$.

The total charge $Q_R$ on the right edge of the system can be found by summing over $k_y$ the effective 1D edge charges $Q_R^{\textrm{1D}}(k_y)$ as given by the 1D surface charge theorem given in Eq.~(\ref{1dsurfcharge}). Using this, the surface charge density on the right edge of the system can be written as
\begin{equation}
    \sigma_R = \frac{Q_R}{aN_y} = \frac{1}{aN_y}\sum_{i=0}^{N_y-1} Q_R^{\textrm{1D}}(k_{yi}),
    \label{surfchargebasic}
\end{equation}
where $N_y$ is the number of lattice sites in the $y$-direction, and $k_{yi} = 2\pi i/aN_y$ for $i = 0,...,N_y-1$ are the discrete admissible $k_y$ values for our finite system.

Let us now consider the flow of the 1D Wannier center during a full cycle of $k_y$, 
which is illustrated in Fig.~\ref{wannier_winding}. There are two possibilities which preserve periodicity in $k_y$. We can have the trivial case where the Wannier center returns back to its original starting position, or the topological case (Chern insulator) where it winds non-trivially, ending up in its starting position but within a different unit cell. In either case, the Wannier center may cross the unit cell boundary at various $k_y$; each time that this occurs, the first term in (\ref{1dsurfcharge}), which lives in a fixed ``home" unit cell, jumps by an integer. Such a jump is generically cancelled out by an equal and opposite jump in the number of charge centers to the right of the home unit cell (given by the second term in (\ref{1dsurfcharge})).

For the trivial case, the above means that the overall 1D edge charges $Q_R^{\textrm{1D}}(k_y)$ do not have any jumps. Indeed, a trivial insulator is characterized by the absence of metallic edge states - any edge states present are gapped from the bulk and so can be assumed to be either fully filled or fully empty, resulting in zero charge transport between the edges. On the other hand, for the topological case, charge \textit{must} be transported from one edge to the other. This is to counteract the bulk Wannier center winding while preserving $k_y$ periodicity.
This corresponds to a Chern insulator which has chiral metallic edge states, whose filling can be made to switch from left to right at arbitrary $k_y$ by altering the chemical potentials on the two edges in the cylindrical geometry.

Since the bulk is translationally invariant in the $x$-direction, the 1D Wannier functions $\ket{w_0(k_y)}$ for the home unit cell in (\ref{1dsurfcharge}) are given by the so-called bulk Hybrid Wannier Functions (HWFs), which for the home unit cell are defined by
\begin{equation}
    \ket{w_{nk_y}} = \sqrt{\frac{a}{2\pi}}\int_0^{2\pi/a}dk_x\ \ket{\psi_{n\textbf{k}}},
\end{equation}
where the $\ket{\psi_{n\textbf{k}}}$'s are the Bloch states for the $n$th band and are normalized according to $\braket{\psi_{n\textbf{k}}}{\psi_{m\textbf{k}'}} = \delta_{nm}\delta(\textbf{k} - \textbf{k}')$. Using (\ref{1dsurfcharge}) with these HWFs in (\ref{surfchargebasic}) and approximating the sum by an integral in the limit of large $N_y$, we can write, to leading order in $N_y$,
\begin{equation} \label{leadingorder}
\begin{aligned}
    \sigma_R = -\frac{e}{2\pi}\int_0^{2\pi/a}dk_y \bigg[\frac{1}{a}\bra{w_{nk_{y}}}\hat{x}\ket{w_{nk_{y}}} + &\mathcal{N}_{k_{y}}\bigg] \\&+ o(1),
\end{aligned}
\end{equation}
where $\mathcal{N}_{k_y}$ is a single integer representing the second term in (\ref{1dsurfcharge}). 

In the trivial case, it is possible to choose a globally smooth gauge for the Bloch states which makes the bulk HWF centers smooth in $k_y$. Since we argued that the overall bracketed expression in (\ref{leadingorder}) does not have any jumps in the trivial case, this makes the integer function $\mathcal{N}_{k_y}$ also smooth and thus constant. As a result, (\ref{leadingorder}) reduces to the standard surface charge theorem (\ref{scthm}) in the thermodynamic limit for the trivial case.  On the other hand, in the topological case, such a globally smooth gauge is not possible. For this case, the bracketed expression will experience a jump as a function of $k_y$ due to the winding of the Wannier centers.

\section{Polarization for Chern insulators}
In this section, we present the main result of this article. We first discuss the difficulties of using surface charge to define polarization for a Chern insulator (namely that it depends on filling of edge states), and ask the question: ``How can polarization be expressed as a bulk quantity for a Chern insulator?" We then show that fractional charge bound to a lattice dislocation provides us with such a bulk polarization.

\subsection{Polarization through surface charge?}
In the previous section, we discussed how a notion of absolute polarization can be defined via the surface charge for a trivial insulator, where both edge and bulk states are insulating. We also discussed some of the differences one encounters when following the same arguments for a Chern insulator. It is therefore natural to ask whether it is still possible to reduce (\ref{leadingorder}) to the form of (\ref{scthm}) in order to obtain a gauge invariant bulk polarization $\textbf{P}$.

For a Chern insulator, we have shown that the bracketed expression in (\ref{leadingorder}) for the surface charge \textit{must} have a discontinuous jump at some $k_y$ due to the non-trivial winding of the Wannier center (which is equivalent to the existence of metallic edge states). Moreover, the precise $k_y$ at which this jump occurs depends on the edge state filling. For instance, one can adjust the chemical potential on one edge of the cylinder, which will certainly alter the resulting surface charge. Indeed, it was shown in \cite{coh09} that while it is possible to write the surface charge for a Chern insulator in the form (\ref{scthm}), the resulting $\textbf{P}$ depends on a choice of momentum space origin arising from the edge chemical potential. As a consequence, the resulting polarization is not a bulk quantity and is not gauge invariant.

These difficulties for Chern insulators can be manifestly seen in the standard Modern Theory of Polarization expressions (\ref{Pe}) and (\ref{wc}). In particular, for a Chern insulator, the Wannier center depends on the chosen gauge for the Bloch states. This is in contrast to a trivial insulator, for which the Wannier center is gauge invariant modulo a real-space lattice vector as long as we consider exponentially localized Wannier functions. In \cite{ourpaper}, it was shown that the Wannier center can be continuously shifted by performing a singular gauge transformation, while maintaining optimal power law decay (which is $\sim 1/r^2$ for a Chern insulator). 

Having polarization as a universal bulk material property is certainly desirable. In the following, we will therefore abandon the present approach of defining polarization through surface charge and instead focus on crystalline defects.

\subsection{Fractional Dislocation Charge} 
A dislocation in a 2D insulator is a topological point defect characterized by a non-trivial translational holonomy which is given by the \textit{Burgers vector} $\textbf{B}$. The role of absolute polarization in the existence of fractional bound charges at dislocations in Topological Crystalline Insulators (TCIs) has been studied extensively in recent times~\cite{vanmiert18, benalcazar19, barkeshli23, zhang2024}. For crystalline insulators that admit an exponentially localized Wannier representation, it has been shown that the dislocation bound charge is given by 
\begin{equation}
    Q_{\textrm{defect}} = \textbf{P}\times\textbf{B} \quad (\textrm{mod}\ e),
    \label{p_cross_b}
\end{equation}
where $\textbf{P}$ is the usual polarization in terms of bulk Wannier centers. For more general TCIs, including when there is no exponentially localized Wannier representation, it has been shown that the dislocation charge can be found in terms of so-called weak topological invariants defined in momentum space, which arise from the underlying crystalline symmetries \cite{ran09, teo10, juricic12, vanmiert18, schindler22}.

In this work, we obtain the dislocation charge for a 2D Chern insulator in terms of a Wannier representation. We show how this provides a natural definition of absolute polarization, by writing the defect charge in the form (\ref{p_cross_b}) and showing that we obtain $\textbf{P}$ as a bulk quantity. Crucially, we do not require any crystalline symmetries beyond translations for this definition. Consequently, our approach is more general than previous results, which either assumed exponentially-localized Wannier states or crystalline symmetries such as rotations or inversion.

In what follows, we focus on a single band with $C = \pm 1$ for simplicity. In Ref.~\cite{ourpaper}, an expression for the \textit{change in polarization} for a 2D Chern insulating band under an adiabatic deformation was found. This work used a gauge for the topological band with a vortex in the Brillouin zone. This allows  the Brillouin zone to be covered by a single, albeit singular, gauge. (See Appendix \ref{app_A} for more information on the vortex construction.)
Using this construction, 
it was shown that the \textit{change in polarization} for a 2D Chern insulating band under an adiabatic deformation can be found in terms of a Wannier representation as
\begin{equation}
       \Delta\textbf{P}_n = -\frac{e }{V_c} \Delta\overline{\textbf{r}}_n + \frac{eC}{2\pi}\hat{\textbf{z}}\times\Delta\textbf{k}_{vn},
    \label{chernpol}
\end{equation}
where $\Delta\overline{\textbf{r}}_n$ is the change in the Wannier center while $\Delta\textbf{k}_{vn}$ is the change in the vortex position of the Bloch states. This requires a choice of gauge where the singularity appears in the form of a vortex, but note that (\ref{chernpol}) extends easily to gauges with multiple vortices, where there is a corresponding term for each vortex. This gauge invariant expression was derived by starting with the adiabatic current and so it provided a physical and unambiguous measure for changes in polarization. 

It is then natural to ask whether we can simply ``cancel out" the $\Delta$'s in (\ref{chernpol}) in order to obtain a notion of absolute polarization. Such a definition would be gauge invariant up to a polarization quantum in the usual way, which is promising (see Appendix \ref{app_C}). It turns out that this is indeed the right definition and we present a direct derivation of this result by carefully computing the dislocation bound charge.

\begin{figure}[t]
    \centering
    \includegraphics[width = \columnwidth]{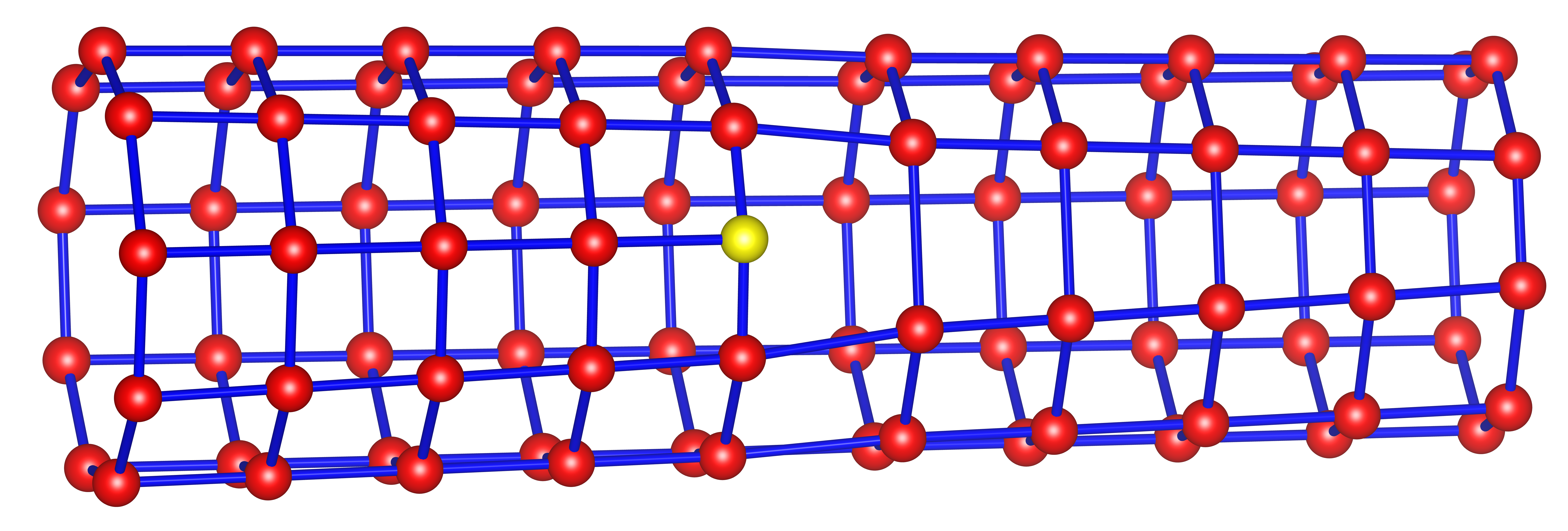}
    \caption{\textbf{Dislocation in a square lattice with a cylindrical geometry.} Note that the right hand edge has one fewer lattice site compared to the left hand edge. The core of the dislocation is the lattice site colored yellow, which has only three nearest neighbors.}
    \label{dislocation_lattice}
\end{figure}

Let us consider the same finite square lattice as in the previous section and create a single dislocation by removing a partial row of atoms at fixed $y$-coordinate. Such a dislocation has a Burgers vector of $\textbf{B} = a(0,1)$. Then, we can put the entire defect lattice into a cylindrical geometry, such that $k_y$ is a good quantum number far away from the dislocation, as shown in Fig.~\ref{dislocation_lattice}. In particular, we can consider the edges of the cylinder to still be translationally invariant in the $y$-direction. We note that the left edge has $N_y$ sites as usual, but the right edge now has only $(N_y-1)$ sites.

We now wish to write down expressions for the total charges on the two edges of this system using the 1D surface charge theorem result (\ref{1dsurfcharge}). Even though the entire system is no longer fully translationally invariant in the bulk due to the defect, we can separate the system into three regions by partitioning out a sufficiently large region centered at the defect. Then, the leftmost region is translationally invariant in its bulk and so we can write the total charge on the left edge of the system in terms of the corresponding bulk HWFs of the left region by summing the effective 1D edge charges (\ref{1dsurfcharge}) over the allowed values of $k_y$ (in the same way that we obtained the 2D surface charge (\ref{surfchargebasic})). This gives, for the left edge,
\begin{equation}
    Q_L = +\frac{e}{a}\sum_{i = 0}^{N_y-1} \bra{w_{nk_{yi}}}\hat{x}\ket{w_{nk_{yi}}}\mode ,
    \label{Totalleftcharge}
\end{equation}
where $k_{yi} = 2\pi i/aN_y$ for $i = 0,...,N_y-1$ as before. Note that the change in sign is due to the fact that we are considering the left-hand edge. Note also that the second term in (\ref{1dsurfcharge}), which contains details about the filling of the metallic edge states, does not appear in (\ref{Totalleftcharge}) because that term is always an integer multiple of $e$. This shows that (\ref{Totalleftcharge}) should be independent of edge filling (since it is taken modulo $e$), which is promising. 

A similar expression can be obtained for $Q_R$, the charge on the right edge, by changing the sign and replacing $N_y$ with $(N_y-1)$ everywhere, noting that this would also lead to a different set of discrete $k_{y}$ values. This discrepancy in the number of sites on the two edges may lead to a fractional contribution to the total edge charge. Crucially, this fractional contribution must be matched by an exactly opposite fractional charge at the point defect, since the total charge of the entire system must be an integer and the defect is the only point in the bulk where translational invariance is broken. In other words, the fractional charge bound to the dislocation must be given by~\cite{vanmiert18}
\begin{equation}
    Q_{\textrm{defect}} = -(Q_L + Q_R)\mode.
    \label{defectcharge1}
\end{equation}
We observe that a dislocation is zero-dimensional, as opposed to an edge which is one-dimensional, and so localizes any fractional charge to a single point. This makes it more natural for defining polarization as opposed to the edge charge density, whose fractional part can change non-trivially even due to integer contributions to the total edge charge.

We are interested in the fractional defect charge (\ref{defectcharge1}) in the thermodynamic limit $N_y \rightarrow \infty$. The edge charge (\ref{Totalleftcharge}) admits a leading order approximation similar to (\ref{leadingorder}), but multiplying an extra factor of $N_y$. Then, if we naively use just the leading order terms, we end up with the defect charge being simply given by $Q_{\textrm{defect}} = -ae(N_y-(N_y-1))\overline{x}_n/V_c \mode = -ae\overline{x}_n/V_c \mode$, which would take us back to a gauge dependent definition of absolute polarization.

However, this cannot be the correct fractional defect charge since we argued above that the fractional part of $Q_L$  given by (\ref{Totalleftcharge}) (and similarly for $Q_R$) must be independent of edge filling. It turns out that we are missing a key term by stopping at leading order. Recall from section \ref{surfcharge} that the bulk HWF centers wind non-trivially for a Chern insulator and that this causes the home unit cell HWF center to jump at some $k_y$. It can be shown that the location of this jump depends on the gauge chosen for the Bloch states. More specifically, the Zak phase, which gives the home unit cell HWF center, jumps precisely at $k_{vy}$, the $k_y$-coordinate of the vortex position of the Bloch states. The reader is referred to Appendix \ref{app_B} for a full derivation of this result. Due to this, it turns out that there are sub-leading order contributions to the edge charges $Q_L$ and $Q_R$ that lead to a non-trivial correction to $Q_{\textrm{defect}}$ in the thermodynamic limit. 

Going to sub-leading order in $N_y$ for the asymptotic expansions of the total edge charges involves approximating the error between the finite sum in (\ref{Totalleftcharge}) and its corresponding integral $(e/a)(aN_y/2\pi)\int_0^{2\pi/a}dk_y\ \bra{w_{nk_y}}\hat{x}\ket{w_{nk_y}} = (eN_y/a)\Bar{x}_n$ in the thermodynamic limit. Note that due to periodicity of the Zak phase and the fact that it has a jump discontinuity, we can write it as the sum of a smooth and periodic function $f_n$ and a sawtooth function. That is,
\begin{equation}
    \frac{1}{a}\bra{w_{nk_{y}}}\hat{x}\ket{w_{nk_{y}}} = f_n(k_y) + C\left(\Theta(k_y - k_{vyn}) - \frac{ak_y}{2\pi}\right),
    \label{decomposition}
\end{equation}
for $k_y \in [0, 2\pi/a]$, where $\Theta$ is the Heaviside step function.

Within the aforementioned error term, we use (\ref{decomposition}) for the quantities in the sum and integral. The contribution from $f_n(k_y)$ can be neglected when $N_y \rightarrow\infty$; this follows from the interesting fact that the error in approximating a finite Riemann sum of $N_y$ points of a smooth, periodic function with the corresponding integral is exponentially small in $N_y$ \cite{rabinowitz}. The contribution from the sawtooth function can be evaluated directly using the mathematical identities
\begin{equation}
    \sum_{m=0}^{M-1} \left(\Theta(m\Delta x - \tilde{x}) - \frac{m}{M}\right) = \frac{M-1}{2} - \left\lfloor\frac{\tilde{x}}{\Delta x}\right\rfloor
\end{equation}
and
\begin{equation}
    \int_0^{M\Delta x}dx\ \left(\Theta(x - \tilde{x}) - \frac{x}{M\Delta x}\right) = \frac{M\Delta x}{2} - \tilde{x},
\end{equation}
where $\lfloor\cdot\rfloor$ is the floor function. This gives, to \textit{sub-leading} order,
\begin{align}
    Q_L &= \frac{aeN_y}{V_c}\overline{x}_n + eC\left(\frac{aN_y}{2\pi}k_{vyn} - \left\lfloor\frac{aN_y}{2\pi}k_{vyn}\right\rfloor - \frac{1}{2}\right) \nonumber \\
    &=  \frac{aeN_y}{V_c}\overline{x}_n + eC\left(\frac{aN_y}{2\pi}k_{vyn} - \frac{1}{2}\right) \mode.
    \label{subleadingorder}
\end{align}
We note that this sub-leading order term is an $\mathcal{O}(1)$ correction to the total charge, and so would go into the $\mathcal{O}(N_y^{-1})$ part of the asymptotic expansion for the surface charge (\ref{leadingorder}). Thus, it vanishes in the thermodynamic limit for the surface charge. On the other hand, it persists as an $\mathcal{O}(1)$ contribution to the fractional defect charge as we shall now see.

Using the same arguments, a similar expression to (\ref{subleadingorder}) can be found for $Q_R$. This is given by
\begin{equation}
    Q_R = -\frac{ae(N_y-1)}{V_c}\overline{x}_n - eC\left(\frac{a(N_y-1)}{2\pi}k_{vyn} - \frac{1}{2}\right) \mode.
\end{equation}
Combining these, we can find the fractional charge bound to the dislocation using (\ref{defectcharge1}), which gives
\begin{equation}
    Q_{\textrm{defect}} = -a\left(\frac{e}{V_c}\overline{x}_n + \frac{eC}{2\pi}k_{vyn}\right) \mode.
    \label{defectcharge2}
\end{equation}
As stated above, we can see that the sub-leading order terms for $Q_L$ and $Q_R$ have contributed an $\mathcal{O}(1)$ correction to the fractional defect charge.
Recalling our choice of $\textbf{B} = a(0,1)$, we can now observe that this is of the form $Q_{\textrm{defect}} = \textbf{P}_n\times\textbf{B} \mode$, where
\begin{equation}
    \textbf{P}_n = -\frac{e}{V_c}\overline{\textbf{r}}_n + \frac{eC}{2\pi}\hat{\textbf{z}}\times\textbf{k}_{vn} \quad \left(\textrm{mod}\ \frac{e\textbf{R}}{V_c}\right).
    \label{abspol}
\end{equation}
Since we considered a Burgers vector of $\textbf{B} = a(0,1)$, (\ref{defectcharge2}) only gives us the $x$-component of $\textbf{P}_n$. However, we can similarly consider a dislocation with $\textbf{B} = a(1,0)$ in order to obtain the $y$-component of $\textbf{P}_n$. Note that this expression for $\textbf{P}_n$ agrees with the formula for relative polarization given by (\ref{chernpol}).

Thus, we can now \textit{define} the absolute polarization of a Chern insulating band via (\ref{abspol}). This is natural for two reasons. Firstly, it straightforwardly extends the polarization-induced defect charge response to a Chern insulator as seen above. Secondly, this definition is \textit{gauge invariant} in the usual way (see Appendix \ref{app_C}),  thus solving the problem we had with using the surface charge to define absolute polarization. In the next section, we demonstrate this result numerically for a specific model.

\section{Numerical Results}
In order to verify our result numerically, we use a spin-rotated Qi-Wu-Zhang (QWZ) model \cite{qwz, ourpaper} with bulk momentum space Hamiltonian
\begin{equation}
H(\textbf{k}) = -\textbf{b}(\textbf{k})\cdot\bm{\sigma},
\label{qwz}
\end{equation}
where $\textbf{b}(\textbf{k})= (u + \cos(k_x) + \cos(k_y),\, \sin(k_x),\, \sin(k_y) + \delta)$. Here $u$ is a parameter, $\delta$ is an onsite staggering potential and $\bm{\sigma} = (\sigma_x, \sigma_y, \sigma_z)$ are the Pauli matrices. We will use $u = 0.5$ for which the $\delta = 0$ model has Chern number $C = 1$. Moreover, the staggering $\delta$ has been introduced to break inversion symmetry and induce a change in the polarization of the model. The corresponding real space model lives on a square lattice with two degrees of freedom per unit cell.

We can then evaluate our formula for the absolute polarization, given by (\ref{abspol}), of this model for a range of values of $\delta$. We note that the gauge invariance of (\ref{abspol})
affords an amount of flexibility in doing so.
We recall that the Wannier center can be computed as an integral of the Berry connection over the Brillouin zone. Furthermore, the location of vortices can be found by numerically computing the circulations over an array of regions covering the Brillouin zone. Building on previous work \cite{ourpaper} we have opted to fix the gauge to correspond to the optimally localized Wannier functions. That said, it should be emphasized that the expression being evaluated is gauge invariant (\ref{abspol}) (up to polarization quanta). Therefore, there are easier ways to compute this quantity in practice. Namely, one can compute 1D Berry phases for each $k_y$ (i.e Zak phases) in a gauge independent fashion (with a fixed choice of branch cut throughout), in which case there will be a jump at some $k_y$. This would correspond to $k_{vyn}$ in our formalism (see Appendix \ref{app_B}). Here we compute $\textbf{P}$ using the first approach in order to explicitly verify (\ref{abspol}) as directly as possible via the construction of optimally localized Wannier functions.

\begin{figure}[t]
    \centering
    \includegraphics[width = \columnwidth]{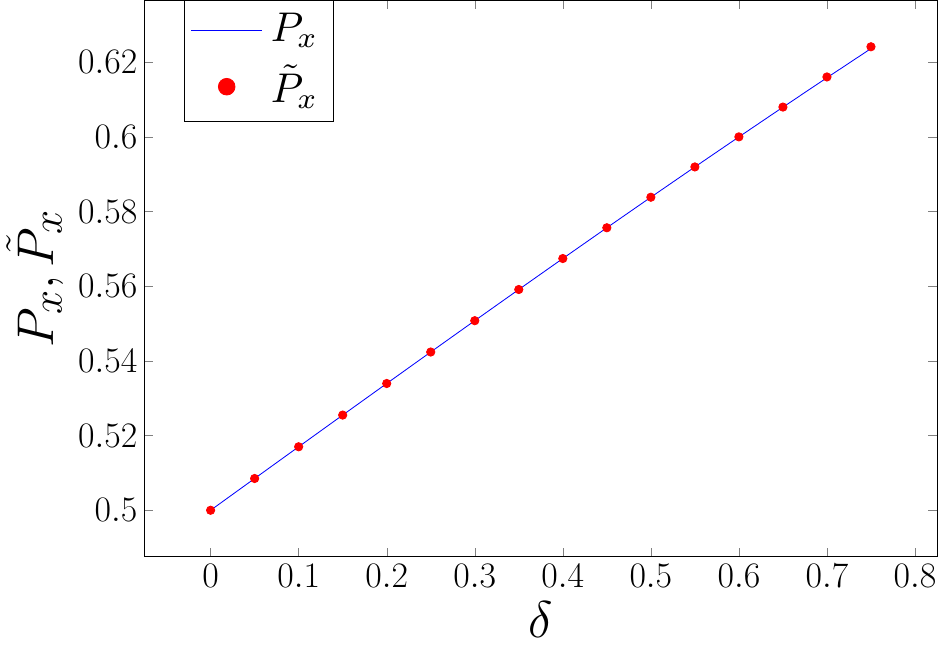}
    \caption{{\bf Plot comparing the theoretical absolute polarization for the QWZ model (blue line) with the numerical value obtained from a real space dislocation charge computation with fully open boundary conditions (red dots).} Note that $P_x$ and $\tilde{P}_x$ have units of $-e/a$. The theoretical values were calculated by constructing Wannier functions and using (\ref{abspol}). The real space dislocation charge calculation was done using clean system sizes ranging from approximately $70a\times 140a$ up to $100a\times 200a$, with charge counted in $R\times R$ regions around the defect for $R = 17a$ and $R = 25a$ respectively.}
    \label{abspolnumerics}
\end{figure}

We then compare our theoretical absolute polarization computed this way with an unbiased real-space calculation of the fractional charge bound to a lattice dislocation in the same model. For this, we inverse Fourier transform our model (\ref{qwz}) into a real-space tight binding model and place \textit{open boundary conditions} on both directions. We then create a dislocation with Burgers vector $\textbf{B} = a(0,1)$ by removing a partial row of atoms and appropriately adjusting the bonds in the real-space Hamiltonian such that the model maintains local translational invariance except at a single site (the ``core" of the dislocation). We then diagonalize this defect Hamiltonian and construct the projector $P$ which projects on to the set of occupied states. Since the bulk unit cell of the system is charge neutral, we can find the fractional charge bound to the defect via
\begin{equation}
    Q_{\textrm{defect}} = -e\textrm{Tr}(PS_{\textrm{defect}}),
\end{equation}
where $S_{\textrm{defect}}$ is a spatial projector which projects onto a sufficiently large region centered around the defect core. This is the same concept used in (\ref{1dcharge}) for the 1D edge charge. Note again that this only works when the electron orbitals are located at the atomic sites. If not, we once again require a suitably smoothened charge distribution. Practically, we choose this region around the defect such that its boundary is equally as far away from the defect as it is from the edge of the system. We emphasize that the charge counted this way is not affected by the filling of edge states since the edges are far away from the region around the defect considered.

The results obtained are displayed in Fig.~\ref{abspolnumerics}. The real space calculation agrees almost perfectly with the theoretical value predicted by (\ref{abspol}), with absolute errors generally ranging from $10^{-6} - 10^{-7}$. The largest relative error is $0.08\%$ for $\delta = 0.75$ and this is because the system size used in the real space calculation is not sufficiently large in comparison to the correlation length for complete convergence in that case. We note that at $\delta = \sqrt{3}/2 \approx 0.866$, a gap closing has occurred and the system is no longer a Chern insulator, and so we do not cross this value. As a final remark, we note that our model has inversion symmetry for $\delta = 0$ and this is correctly reflected in the quantized polarization of $0.5$ obtained in this case. For this case, the real space model has a filling anomaly~\cite{benalcazar19} which must be taken into account, meaning that we cannot have exactly half filling due to a two-fold degeneracy at zero energy, with one state localized to the defect and the other to the edge. Therefore, we must fill either both or neither of these states in order to correctly count the charge around the defect up to integer. These bound states are a by-product of the non-trivial polarization and the crystalline symmetry in the clean $\delta = 0$ model and the existence of a defect. We do not observe this filling anomaly in the clean real space model as it is concealed by the metallic edge states (unlike for an obstructed atomic insulator).

\section{Conclusion}
In this article, we have shown that a microscopic definition of absolute polarization can be obtained unambiguously for a 2D Chern insulator by using fractional charges at lattice dislocations. This was done without assuming any crystalline symmetries apart from translations, thus generalizing existing work. Our result is consistent with recent work for TCIs, such as Ref. \cite{Vaidya24} which studied differences in polarization via interface charge, and Ref. \cite{barkeshli23}, which studied absolute polarization via topological field theory. Finally, we demonstrated our result numerically by considering a particular tight-binding Chern insulator model.
We approached the polarization from the perspective of bound charge on a crystalline defect. 
An interesting future direction will be to answer the question of whether our definition of polarization correctly predicts other bulk effects such as the electric field experienced by a free electron placed into the system. We conjecture that it should, since it is the only way to make the usual Wannier center definition gauge invariant.

\begin{acknowledgments}
We are grateful to Andres Perez Fadon for discussions and helpful comments. T.M.G acknowledges support from an Imperial Department of Mathematics Roth PhD studentship.
F.S. was supported by a UKRI Future Leaders Fellowship MR/Y017331/1.
R.B. acknowledges support from an
Imperial Mathematics Research Impulse Grant and hospitality of the Aspen
Center for Physics under NSF Grant No. PHY-2210452.
\end{acknowledgments}

\appendix

\section{Vortices in Bloch states for a Chern insulator}
\label{app_A}
In this section we briefly study Bloch state vortices for a Chern insulator. As discussed in the main text, it is impossible to pick a globally smooth gauge for the Bloch states of a Chern insulating band. Instead, we can construct a gauge for which the Bloch states are smooth everywhere except at a single point in the Brillouin zone. The nature of this singularity is characterized by the Chern number, which is the integer topological invariant given by
\begin{equation}
    C = \frac{1}{2\pi}\intbz\ \Omega(\textbf{k}),
\end{equation}
where $\Omega(\textbf{k})$ is the smooth, gauge invariant Berry curvature. Note that $\Omega(\textbf{k}) = \nabla_{\textbf{k}}\times\textbf{A}(\textbf{k})$ in regions of the Brillouin zone where the Bloch states are smooth. However, it is crucial to note that this is \textit{not} true at the singularity. Instead, since the Berry connection is periodic across the Brillouin zone, we can use Stokes' theorem to argue that \cite{ourpaper}
\begin{equation}
    \nabla_{\textbf{k}}\times\textbf{A}(\textbf{k}) = \Omega(\textbf{k}) - 2\pi C\delta_P(\textbf{k} - \textbf{k}_v),
    \label{keyeqn}
\end{equation}
where $\delta_P$ is the Brillouin zone periodic Dirac delta function and $\textbf{k}_v$ is the position of the singularity. Note that the integral of the RHS of the above over the entire Brillouin zone is zero, which is consistent with Stoke's theorem. We then refer to the singularity at $\textbf{k}_v$ as a vortex because the phase of the Bloch states winds non-trivially around it. To see this, note that we can use a gauge transformation to make the Bloch states \textit{locally smooth} around $\textbf{k}_v$. In other words, inside a disc $D_{\epsilon}$ of sufficiently small radius $\epsilon > 0$ about $\textbf{k}_v$, we can write
\begin{equation}
    \ket{u_{\textbf{k}}} = e^{-i\theta(\textbf{k} - \textbf{k}_v)}\ket{\tilde{u}_{\textbf{k}}},
\end{equation}
where $\ket{\tilde{u}_{\textbf{k}}}$ is smooth inside $D_{\epsilon}$. Then, if we integrate $\nabla_{\textbf{k}}\times\textbf{A}(\textbf{k})$ over the entire Brillouin zone while noting that $\nabla_{\textbf{k}}\times\textbf{A}(\textbf{k}) = \nabla_{\textbf{k}}\times\tilde{\textbf{A}}(\textbf{k}) + \nabla_{\textbf{k}}\times\nabla_{\textbf{k}}\theta(\textbf{k} - \textbf{k}_v)$ in $D_{\epsilon}$, we find that
\begin{equation}
    \int_{D_{\epsilon}}d\textbf{k}\ \nabla_{\textbf{k}}\times\nabla_{\textbf{k}}\theta(\textbf{k} - \textbf{k}_v) = -2\pi C.
\end{equation}
Thus, letting $\epsilon \rightarrow 0$, we deduce that
\begin{equation}
    \nabla_{\textbf{k}}\times\nabla_{\textbf{k}}\theta(\textbf{k}-\textbf{k}_v) = -2\pi C\delta_P(\textbf{k} - \textbf{k}_v),
\end{equation}
and so we say that the Bloch states have a vortex of winding number $-C$ at $\textbf{k}_v$. We may later also refer to this as an antivortex because of the relative minus sign with respect to the Chern number.

\section{Jump in the Zak phase}
\label{app_B}
In this section, we study the jump discontinuity in the Zak phase, which gives the HWF centers, for a Chern insulator. The Zak phase \cite{zak} is defined to be the Berry phase acquired by the Bloch states in a path across the Brillouin zone. For a path $\gamma$ across the Brillouin zone, the Zak phase is given by
\begin{equation}
    \phi_{\gamma} = \oint_{\gamma} d\boldsymbol{\ell}\cdot\textbf{A}({\bf k}),
    \label{zak}
\end{equation}
where $\textbf{A}$ is the Berry connection. Note that such a path $\gamma$ is a closed curve due to the periodicity of the Brillouin zone.

\begin{figure}
    \centering
    \includegraphics[width=0.85\columnwidth]{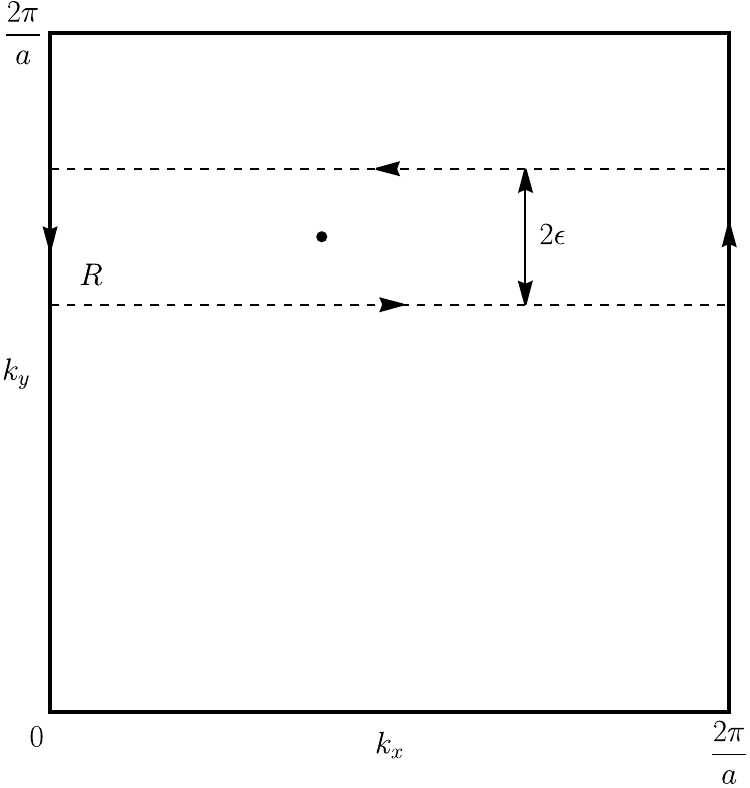}
    \caption{Figure showing the integration region to demonstrate the jump in the Zak phase.}
    \label{zakJump}
\end{figure}

Specializing to our case where we have a square (or rectangular) Brillouin zone, the Zak phase for a horizontal path across the Brillouin zone for a fixed $k_y$ is given by
\begin{equation}
    \phi_x(k_y) = \int_0^{2\pi/a} dk_x\ A_x(\textbf{k}),
    \label{zakh}
\end{equation}
where $a$ is the lattice constant.
Recall from the main text that the hybrid Wannier functions characterized by $k_y$, which is a good quantum number when we have translational invariance in the $y$-direction, are given by
\begin{equation}
    w_{nk_y}(\textbf{r}) = \sqrt{\frac{a}{2\pi}}\int_0^{2\pi/a} dk_x\ \psi_{n\textbf{k}}(\textbf{r}),
    \label{hwf}
\end{equation}
where $\psi_{n\textbf{k}}(\textbf{r})$ are the Bloch states for the $n$th band, and $a$ is the lattice constant. Dropping the band index, the corresponding hybrid Wannier centers can be written in terms of the Zak phase (\ref{zakh}) as
\begin{equation}
    \bra{w_{k_y}}\hat{x}\ket{w_{k_y}} = \frac{a}{2\pi}\phi_x(k_y).
    \label{hwfc}
\end{equation}
Let us now consider a Chern insulator with Chern number $C$. Then, as seen in the previous section, there must be at least one vortex in the Bloch states. We shall consider the case of just a single vortex at position $\textbf{k}_v = (k_{vx}, k_{vy})$, which must have winding number $C$. We wish to understand what happens to the Zak phase $\phi(k_y)$ as $k_y$ passes through $k_{vy}$. Consider a rectangular region $R_{\epsilon}$ of width $2\epsilon$, where $\epsilon > 0$, in the Brillouin Zone with the vortex inside it, as shown in Fig.~\ref{zakJump}.

Then, using Stokes' theorem and \ref{keyeqn}, we have
\begin{equation}
    \oint_{\partial R_{\epsilon}}d\boldsymbol{\ell}\cdot\textbf{A}(\textbf{k}) = \int_{R_{\epsilon}} d\textbf{k}\ \nabla_{\textbf{k}}\times\textbf{A}(\textbf{k}) = \int_{R_{\epsilon}} d\textbf{k}\ \Omega(\textbf{k}) - 2\pi C.
    \label{stokes}
\end{equation}
On the other hand, a direct calculation gives
\begin{widetext}
\begin{equation}
    \oint_{\partial R_{\epsilon}}d\boldsymbol{\ell}\cdot\textbf{A}(\textbf{k}) = \int_0^{2\pi/a}dk_x\ \bigg[A_x(k_x, k_{vy}-\epsilon) - A_x(k_x, k_{vy}+\epsilon)\bigg] + \int_{k_{vy}-\epsilon}^{k_{vy}+\epsilon}dk_y\ \bigg[A_y\left(\frac{2\pi}{a},k_y\right) - A_x(0, k_y)\bigg].
    \label{direct}
\end{equation}
\end{widetext}
Taking the limit of both (\ref{stokes}) and (\ref{direct}) as $\epsilon \rightarrow 0$ yields
\begin{equation}
    \phi_x(k_{vy}^+) - \phi_x(k_{vy}^-) = 2\pi C
\end{equation}
where $\phi_x(k_{vy}^{\pm}) = \lim_{\epsilon \rightarrow 0^{\pm}}\phi_x(k_{vy} + \epsilon)$ are the right and left limits respectively. Thus, the Zak phase jumps by $2\pi C$ when we pass through the vortex. The hybrid Wannier center, given by (\ref{hwfc}), then correspondingly jumps by $aC$ and therefore the 1D polarization jumps by $C$.

\section{Gauge invariance of the absolute polarization}
\label{app_C}
In this section, we show that the expression we obtained for the absolute polarization is gauge invariant. We begin by recalling that the polarization for a Chern insulating band is given by
\begin{equation}
    \textbf{P}_n = -\frac{e}{V_c}\overline{\textbf{r}}_n + \frac{eC}{2\pi}\hat{\textbf{z}}\times\textbf{k}_{vn}\ \ \left(\textrm{mod}\ \frac{e\textbf{R}}{V_c}\right),
    \label{abspolsup}
\end{equation}
where $\overline{\textbf{r}}_n = \frac{1}{V_{\textrm{BZ}}}\intbz\ \textbf{A}(\textbf{k}) $ is the Wannier center and $\textbf{k}_{vn}$ is the vortex position.
Now consider a gauge transformation of the Bloch states of the form $\ket{u_{n\textbf{k}}} \rightarrow e^{-i\theta(\textbf{k})}\ket{u_{n\textbf{k}}}$. Then, the Berry connection transforms as 
\begin{equation}
    \textbf{A}_n(\textbf{k}) \rightarrow \textbf{A}_n(\textbf{k}) + \nabla_{\textbf{k}}\theta(\textbf{k}).
    \label{Atransform}
\end{equation} 
If the position of the vortex is not changed under this gauge transformation, then it is clear that the Wannier center can only move by a lattice vector and so $\textbf{P}_n$ is gauge invariant. Therefore, let us focus on so-called \textit{singular gauge transformations}, which are gauge transformations that move the position of the vortex.

Consider a gauge transformation that moves the vortex from $\textbf{k}_{vn}$ to $\tilde{\textbf{k}}_{vn}$. Then, in light of (\ref{Atransform}) and (\ref{keyeqn}), we must have
\begin{equation}
    \nabla_{\textbf{k}}\times\nabla_{\textbf{k}}\theta(\textbf{k}) = 2\pi C(\delta_P(\textbf{k} - \textbf{k}_{vn}) - \delta_P(\textbf{k} - \tilde{\textbf{k}}_{vn})).
\end{equation}
This can be thought of as creating a vortex-antivortex pair of winding numbers $C$ and $-C$ respectively at positions $\textbf{k}_{vn}$ and $\tilde{\textbf{k}}_{vn}$ respectively.
\begin{figure}
    \centering
    \includegraphics[width=0.85\columnwidth]{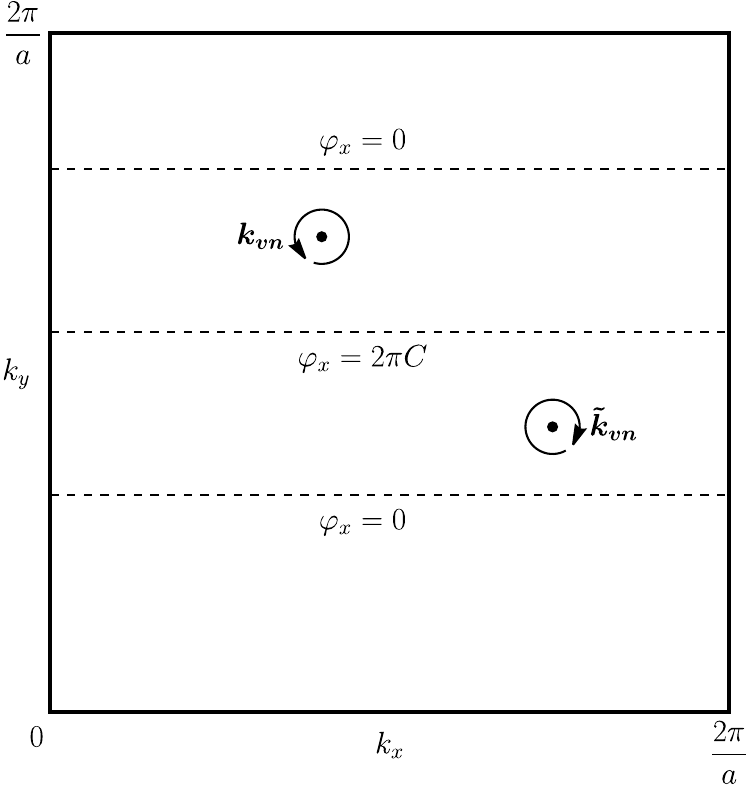}
    \caption{Figure showing values of $\varphi_x(k_y)$ at different values of $k_y$ in the Brillouin zone for a gauge transformation with a vortex-antivortex pair. An anti-clockwise arrow represents a circulation of $2\pi C$.}
    \label{gaugeinvariance}
\end{figure}
Moreover, under such a gauge transformation, $\textbf{P}_n$ transforms as
\begin{align}
    \textbf{P}_n \rightarrow \textbf{P}_n - \frac{e}{(2\pi)^2}&\intbz\ \nabla_{\textbf{k}}\theta(\textbf{k}) \nonumber \\
    &+ \frac{eC}{2\pi}\hat{\textbf{z}}\times\left(\tilde{\textbf{k}}_{vn} - \textbf{k}_{vn}\right)\ \ \left(\textrm{mod}\ \frac{e\textbf{R}}{V_c}\right).
    \label{Ptransform}
\end{align}
Focusing on just the $x$-component, note that
\begin{equation}
    \intbz\ \frac{\partial\theta}{\partial k_x} = \int_0^{2\pi/a}dk_y\ \varphi_x(k_y),
    \label{17}
\end{equation}
where
\begin{equation}
    \varphi_x(k_y) = \int_0^{2\pi/a}dk_x\ \frac{\partial\theta}{\partial k_x} = 2\pi n(k_y),
\end{equation}
for an integer function $n(k_y)$. Note that this follows from the fact that $e^{i\theta}$ is required to be Brillouin zone periodic. We shall focus on a specific class of gauge transformations for which $n(k_y) = 0$ if $\theta$ is smooth. Thus, for singular gauge transformations, which are manifestly not globally smooth, we shall assume that any non-zero values for $n(k_y)$ are caused by the introduction of the vortex-antivortex pair.

Suppose for the moment that $k_{vyn} > \tilde{k}_{vyn}$. Then, using arguments identical to Appendix \ref{app_B} (essentially treating $\varphi_x(k_y)$ as a Zak phase), we obtain that
\begin{equation}
    \varphi_x(k_y) =
    \begin{cases}
        0 & 0\leq k_y<\tilde{k}_{yvn}\\
        2\pi C & \tilde{k}_{yvn}<k_y<k_{yvn} \\
        0 & k_{yvn}<k_y\leq 2\pi/a,
    \end{cases}
\end{equation}
as illustrated in Fig.~\ref{gaugeinvariance}. Thus, using (\ref{17}), we obtain
\begin{equation}
    \intbz\ \frac{\partial\theta}{\partial k_x} = -2\pi C\left(\tilde{k}_{yvn} - k_{yvn}\right).
\end{equation}
It can be shown that the same result holds when $k_{vyn} > \tilde{k}_{yvn}$. Moreover, a similar approach can be used to obtain, for the $y$-component, that
\begin{equation}
    \intbz\ \frac{\partial\theta}{\partial k_y} = 2\pi C\left(\tilde{k}_{xvn} - k_{xvn}\right)
\end{equation}
Thus, we have that
\begin{equation}
    \intbz\ \nabla_{\textbf{k}}\theta(\textbf{k}) = 2\pi C\hat{\textbf{z}}\times\left(\tilde{\textbf{k}}_{vn} - \textbf{k}_{vn}\right).
\end{equation}
Putting everything together, we obtain that
\begin{equation}
    -\frac{e}{(2\pi)^2}\intbz\ \nabla_{\textbf{k}}\theta(\textbf{k}) + \frac{eC}{2\pi}\hat{\textbf{z}}\times\left(\tilde{\textbf{k}}_{vn} - \textbf{k}_{vn}\right) = 0,
\end{equation}
and so returning to (\ref{Ptransform}), we see that $\textbf{P}_n$ is indeed gauge invariant.

As a final remark, we note that a singular gauge transformation cannot be constructed such that it annihilates the original vortex but does not create a new one. We should expect this to be impossible since we would be changing the total winding of the Bloch state vortices, thus changing the Chern number. Indeed, note that if $e^{i\theta(\textbf{k})}$ were periodic, then $\nabla_{\textbf{k}}\theta(\textbf{k})$ would also be periodic and so has zero circulation around the Brillouin zone. Thus, by Stokes' theorem, we must have $\intbz\ \nabla_{\textbf{k}}\times\nabla_{\textbf{k}}\theta(\textbf{k}) = 0$. In other words, the total winding number of any vortices within the gauge transformation has to be zero if it is to be periodic.

\bibliography{biblio}

\end{document}